\documentclass[letter]{aa}  

\usepackage{subcaption}
\usepackage{graphicx}
\usepackage{natbib}
\usepackage{txfonts}
\begin{document} 

   \title{Non-star-forming molecular gas in the Abell 1367 intra-cluster multiphase orphan cloud\thanks{Based on observations carried out under project number 077-21 with the Institut de Radio Astronomie Millimétrique (IRAM) 30-meter telescope. IRAM is supported by INSU/CNRS (France), MPG (Germany) and IGN (Spain).}}

   \titlerunning{Molecular gas in intra-cluster orphan cloud}
   \authorrunning{Pavel J\'achym et al.}

   \author{Pavel J\'achym
          \inst{1},
          Ming Sun
          \inst{2},
          Masafumi Yagi
          \inst{3},
          Chong Ge
          \inst{2},
          Rongxin Luo
          \inst{2},
          Fran\c coise Combes
          \inst{4},
          Ane\v zka Kab\'atov\'a
          \inst{1},
          Jeffrey D.~P. Kenney
          \inst{5},
          Tom C. Scott
          \inst{6},
          \and
          Elias Brinks
          \inst{7}
          }

   \institute{Astronomical Institute of the Czech Academy of Sciences,
              Bo\v{c}n\'{i} II 1401, 141 00, Prague, Czech Republic\\
              \email{jachym@ig.cas.cz}
         \and
             Department of Physics and Astronomy, University of Alabama 
             in Huntsville, Huntsville, AL 35899, USA
         \and
             National Astronomical Observatory of Japan, 2-21-1, Osawa, 
             Mitaka, Tokyo, 181-8588, Japan
         \and
             Observatoire de Paris, LERMA, Coll\`ege de France, CNRS, 
             PSL University, Sorbonne University, 75014 Paris, France
         \and
             Department of Astronomy, Yale University, 46 Hillhouse Avenue, New Haven, CT 06511, USA
         \and
             Institute of Astrophysics and Space Sciences (IA), Rua das Estrelas, P-4150-762 Porto, Portugal
         \and
              Centre for Astrophysics Research, University of Hertfordshire, College Lane, Hatfield AL10 9AB, UK
             }

   \date{Received ...; accepted ...}


  \abstract{
We report the detection of CO emission in the recently discovered multiphase isolated gas cloud in the nearby galaxy cluster Abell 1367. The cloud is located about 800~kpc in projection from the center of the cluster and at a projected distance of $> 80$~kpc from any galaxy. It is the first and the only known isolated ``intra-cluster'' cloud detected in X-ray, H$\alpha$, and CO emission. We found a total of about $2.2\times 10^8~M_\odot$ of H$_2$ with the IRAM 30-m telescope in two regions, one associated with the peak of H$\alpha$ emission and another with the peak of X-ray emission surrounded by weak H$\alpha$ filaments. The velocity of the molecular gas is offset from the underlying H$\alpha$ emission by $> 100$~km\,s$^{-1}$ in the region where the X-ray peaks. The molecular gas may account for about $10 \%$ of the total cloud's mass, which is dominated by the hot X-ray component. The previously measured upper limit on the star formation rate in the cloud indicates that the molecular component is in a non-star-forming state, possibly due to a combination of low density of the gas and the observed level of velocity dispersion. The presence of the three gas phases associated with the cloud suggests that gas phase mixing with the surrounding intra-cluster medium is taking place. The possible origin of the orphan cloud is a late evolutionary stage of a ram pressure stripping event. In contrast, the nearby ram pressure stripped galaxy 2MASX~J11443212+2006238 is in an early phase of stripping and we detected about $2.4\times 10^9~M_\odot$ of H$_2$ in its main body.
}


   \maketitle
%

\section{Introduction}
Abell 1367 (A1367) is a nearby ($z= 0.022$, $D=96$~Mpc) dynamically unrelaxed cluster in the Coma supercluster with $R_{500}= 924$~kpc and at least two subclusters merging along the southeast–northwest direction \citep{ge2019}. The cluster’s total mass is $\sim 3.3\times 10^{14}~M_\odot$ \citep{boselli2006}, which is about one-third of that of the Coma cluster. There are several ram pressure stripped (RPS) galaxies with offset HI distributions and multiphase tails and a prominent infalling group with an associated X-ray and H$\alpha$ trail caused by ram pressure stripping \citep[e.g.,][]{gavazzi2001, gavazzi2003, iglesias-paramo2002, scott2010, scott2018, yagi2017, fossati2019}. Recently, an isolated cloud at about 800~kpc in projection from the center of the cluster, not far from the major axis of the cluster, was discovered in narrow-band H$\alpha$ imaging \citep{yagi2017}. Due to the absence of an obvious parent galaxy within an $80$~kpc search radius, the object was suggested as an orphan cloud (OC). It is projected at the base of a filament of galaxies which extends to the Coma cluster. It is also close to the cluster merger shock. Follow-up XMM observations in the field \citep{ge2019} revealed diffuse soft X-ray emission around the same position as the H$\alpha$ OC. The X-ray OC is asymmetric around its X-ray peak, and a radial surface brightness profile centered on its peak shows an effective radius of $\sim 30$~kpc. VLT/MUSE IFU observations \citep{ge2021} reveal that the H$\alpha$ OC is elongated ($\sim 10$~kpc~$\times\, 20$~kpc) with a set of narrow filaments connected to the main body. There is a velocity gradient of $\sim 200$~km\,s$^{-1}$ in nearly the east-west direction.

\begin{figure*}
  \centering
  \begin{minipage}{0.32\textwidth}
    \includegraphics[width=0.9\textwidth]{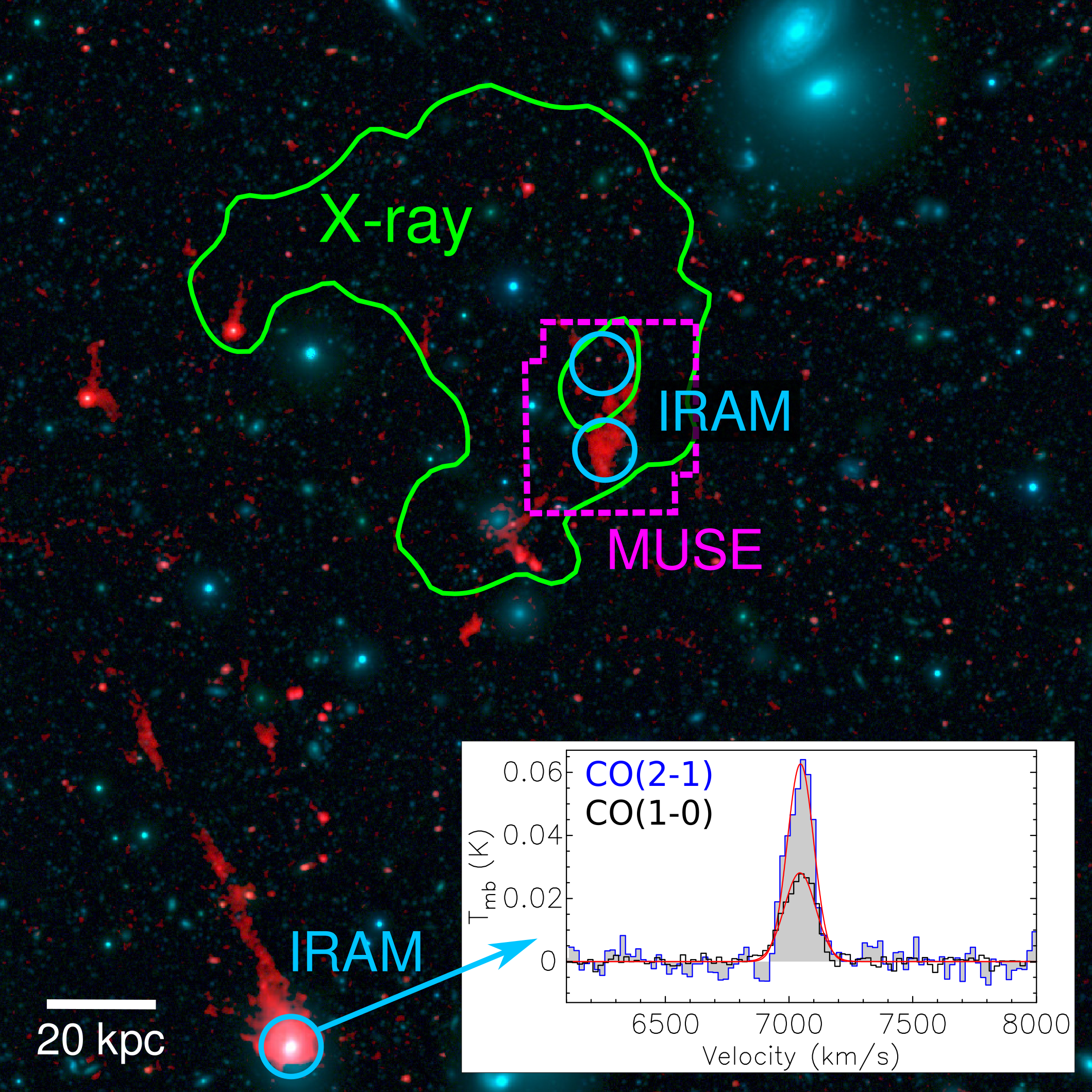}
  \end{minipage}
  \hspace{-0.8cm}
  \begin{minipage}{0.32\textwidth}
    \includegraphics[width=1.1\textwidth]{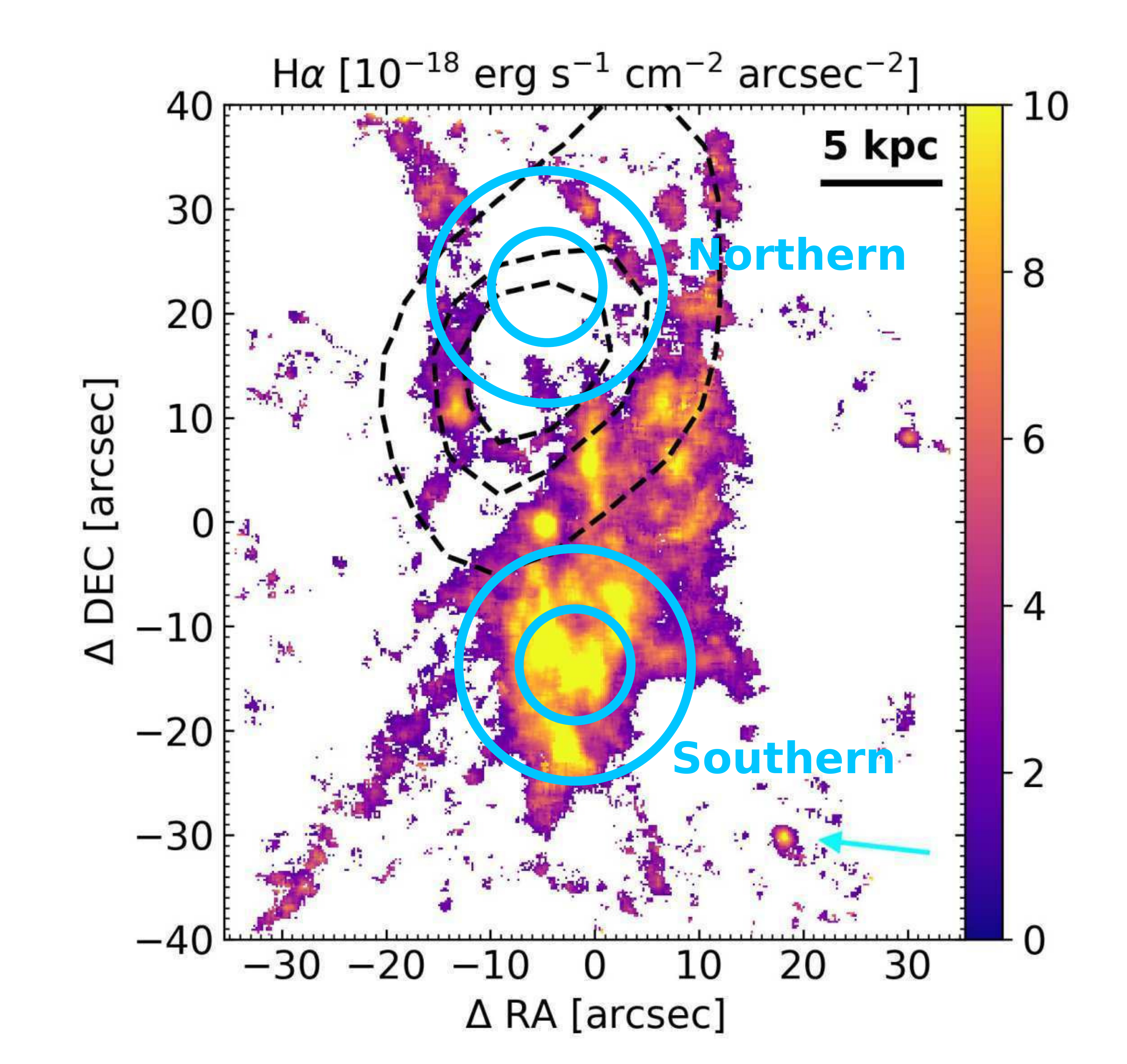}
  \end{minipage}
  \hspace{0.5cm}
  \begin{minipage}{0.32\textwidth}
    \includegraphics[width=0.9\textwidth]{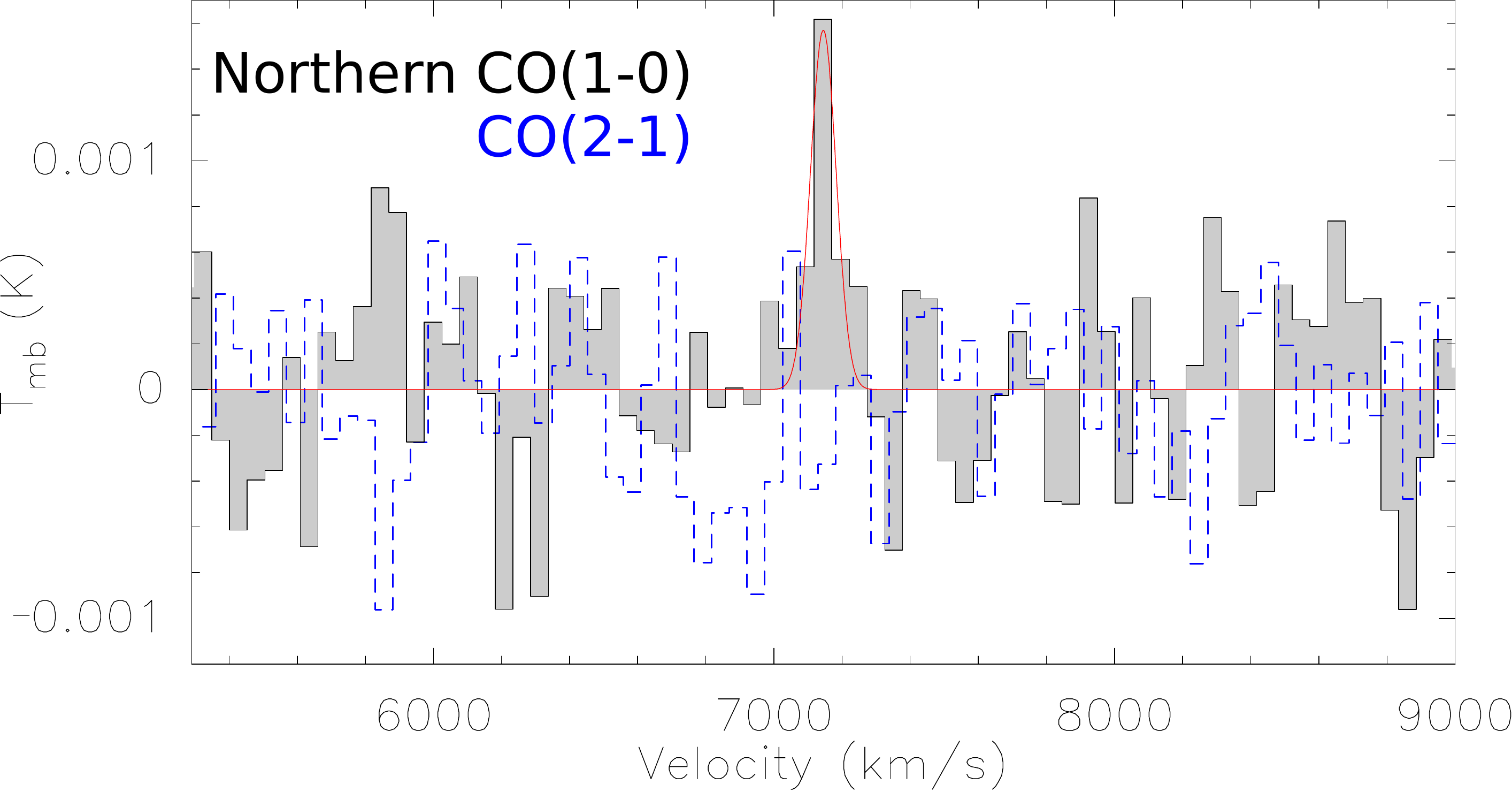}\vspace{0.2cm}
    \includegraphics[width=0.9\textwidth]{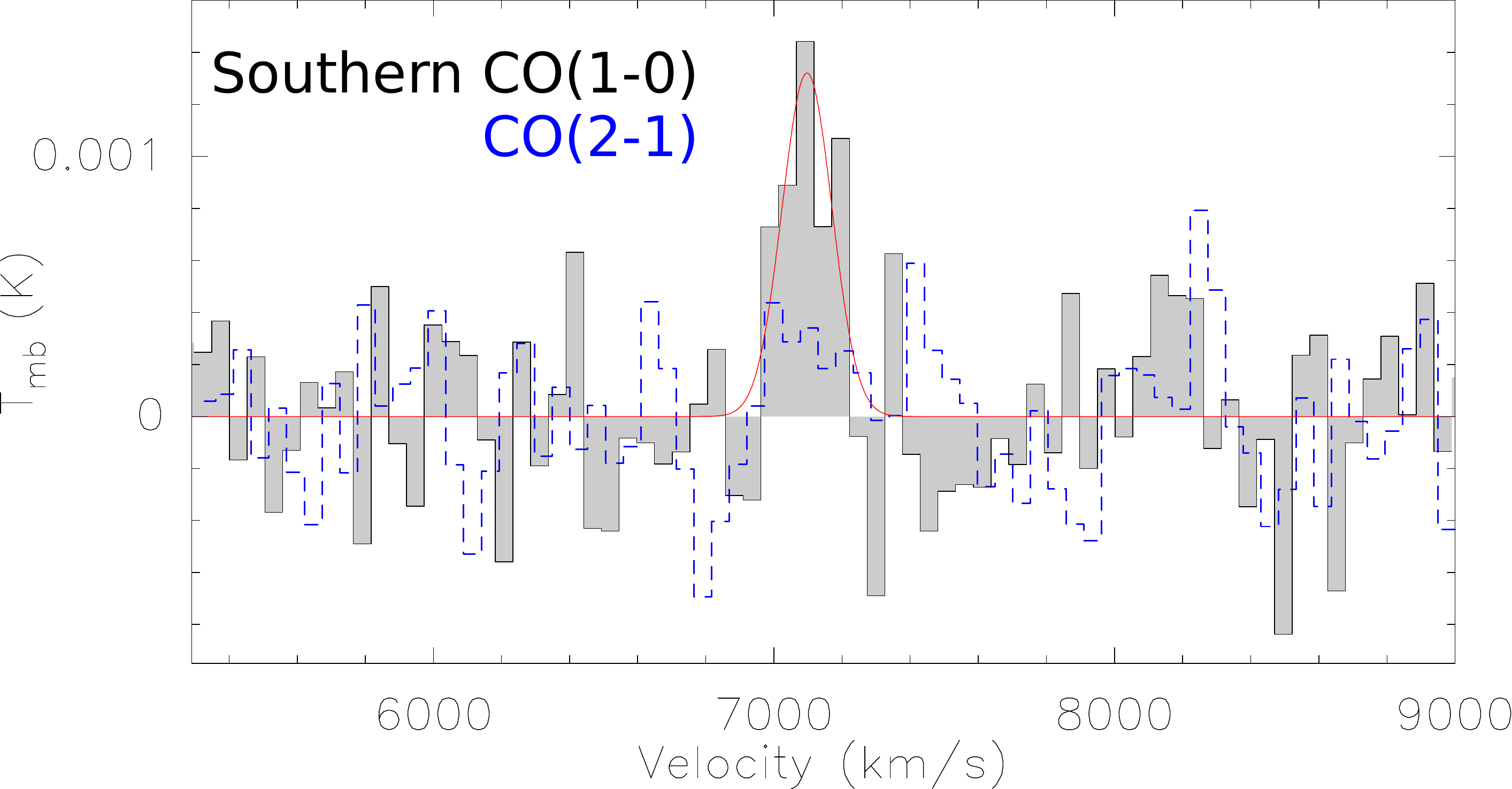}
  \end{minipage}
  \caption{
Multiwavelength observations of the A1367 OC. \textit{Left:} A1367 OC -- the green contour outlines its X-ray morphology ($0.5- 2$~keV {\it XMM} observations) overlaid on a \textit{Subaru} three-color composite image (red, H$\alpha$; green, $r$-band; and blue, $g$-band). The dashed magenta region shows the MUSE FOV, and the blue circles indicate the IRAM CO(1-0) beams. Image adapted from \citet{ge2021}. The galaxy 2MASX~J11443212+2006238 with its 85~kpc long H$\alpha$ ram pressure stripped tail is visible to the south of the OC. The CO spectra from a pointing centered on this galaxy are shown in the inset. \textit{Middle:} MUSE H$\alpha$ image of the OC \citep{ge2021} showing the positions of the CO(1-0) and CO(2-1) beams (blue circles, HPBW of $21.7''\approx 9.7$~kpc and $10.9''\approx 4.9$~kpc, respectively). The black dashed contours show the peak X-ray emission of the OC. \textit{Right:}  CO(1-0) lines detected at the positions over the northern X-ray peak (top panel) and the southern H$\alpha$ peak (bottom panel). Velocity corresponds to $\upsilon_{\rm rad}= cz/(z+1)$. The corresponding CO(2-1) spectra divided by a factor of 3 are shown with dotted blue lines. In the southern region (bottom panel), a marginal CO(2-1) line detection is suggested from a positive signal at velocities around the CO(1-0) line central velocity.
  }
  \label{FigOC}
\end{figure*}

It is likely that the origin of the OC is ISM stripped from an infalling galaxy. Whether it was due to ram pressure stripping, a tidal interaction, or a combination of both, is not clear. Assuming that the speed of the OC to the parent galaxy is less than $\sim 1000$~km\,s$^{-1}$ (cf. the A1367 radial velocity dispersion of 726~km\,s$^{-1}$), the OC should have survived in the cluster environment for $>70$~Myr. The MUSE observations and line diagnostics suggest high, about solar metallicity of the warm ionized gas \citep{ge2021}, implying a massive parent for the OC. The temperature of the X-ray OC, $1.6\pm 0.1$~keV, is higher than typical temperatures of X-ray RPS tails of cluster late-type galaxies \citep[$\sim 0.9$~keV,][]{sun2010}, which, together with the large distance to any possible parent galaxy, suggests an advanced evolutionary stage of the OC as it mixes with the surrounding ICM. The bright H$\alpha$ part of the cloud may be associated with the only surviving cold clouds, while the bulk of the X-ray OC is free of cold gas now. The A1367 OC presents a unique laboratory to study the evolution of the ISM far away from the parent galaxy.

Here we present the first CO observations of the H$\alpha$-bright part of the OC. These observations show that the OC also contains a cold molecular gas component, in addition to the warm and hot ionized gas phases. Despite the presence of the molecular gas, only a very small amount of star formation was found in the OC \citep{ge2021}, indicating that the conversion of the molecular gas to stars is extremely inefficient.
The A1367 OC with its multiphase components of cold, warm, and hot gas is unique. In the Virgo cluster, several isolated HI clouds with no optical counterparts are known \citep{davies2004, kent2007}, as well as the low-mass, star-forming H$\alpha$ cloud SECCO~1 \citep{beccari2017}. There are also examples of isolated HI clouds in galaxy groups \citep[e.g.,][]{wong2021}.

\section{Observations}
The observations were carried out with the IRAM 30-m telescope operated by the Institut de Radio Astronomie Millimétrique (IRAM) at Pico Veleta, Spain, from 01 June 2021 to 07 June 2021 (PI J\'achym, project ID 077-21). The EMIR receiver in the E090 and E230 bands was used to observe at the frequencies of the $^{12}$CO(1-0) ($\nu_{\rm rest} = 115.271$~GHz) and the $^{12}$CO(2-1) ($\nu_{\rm rest} = 230.538$~GHz) lines simultaneously. The FTS spectrometer, together with the WILMA autocorrelator, were connected to both receivers. Symmetric wobbler switching with the maximum secondary reflector throw of $\pm 60~\arcsec$ and $t_{\rm phase}= 1.5$~Hz was used. Observing conditions varied from excellent with PWV as low as $1-2$~mm to bad with $\rm{PWV}>10$~mm. Two nights (04 \& 05 June) were lost due to unstable conditions and high wind. 

Two regions of the OC were observed, one ("southern", J2000 RA 11:44:22.89, Dec. +20:10:31.5) covering the main peak of the H$\alpha$ and the other ("northern", J2000 RA 11:44:23.17, Dec. +20:11:07.6) covering the neighboring peak of the X-ray emission (see Fig.~\ref{FigOC}, middle panel). The latter region lacks bright H$\alpha$ features. These were deep observations, with on-source time for the southern and northern regions of 4.85~hr and 3.8~hr, respectively. In addition, we targeted the nearby galaxy 2MASX~J11443212+2006238 with a pointed observation of 0.8~hr. The half power beamwidth (HPBW) of the IRAM 30m main beam is $\sim 21.7''$ and $10.9''$ at the CO(1-0) and CO(2-1) frequencies, respectively. The data were reduced using CLASS from the GILDAS\footnote{http://www.iram.fr/IRAMFR/GILDAS} software package developed at IRAM. The corrected antenna temperatures, $T_{\rm A}^*$, provided by the IRAM 30m calibration pipeline were converted to a main-beam brightness temperature by $T_{\rm mb}= T_{\rm A}^* F_{\rm eff}/\eta_{\rm mb}$, using a main beam efficiency of $\eta_{\rm mb}= 0.78$ at 115~GHz and 0.59 at 230~GHz, and the forward efficiencies $F_{\rm eff}$ of 0.94 and 0.92, respectively. The flux density-to-main beam antenna temperature conversion factor is $\sim 5$~Jy\,beam$^{-1}$\,K$^{-1}$ for both bands \citep[see][]{jachym2017}.

\begin{table*}
\caption{Properties of the CO detections in the OC and the RPS galaxy 2MASX~J11443212+2006238.}
\label{table:1}
\centering
\begin{tabular}{ccccccccc}
\hline\hline
Region & Line & $\sigma_{\rm rms}$ & $cz/(z+1)$ & FWHM & $T_{\rm mb}^{\rm peak}$ & $I_{\rm CO}$ & $L_{\rm CO}$ & $M_{\rm H_2}$ \\
 & & (mK) & (km\,s$^{-1}$) & (km\,s$^{-1}$) & (mK) & (K\,km\,s$^{-1}$) & ($10^7$\,K\,km\,s$^{-1}$\,pc$^2$) & ($10^8~M_\odot$)\\
\hline
Northern & CO(1-0) & 0.46 & $7145.1\pm 14.3$ & $85.6\pm 30.7$  & 1.55 & $0.14\pm 0.05$ & $1.7\pm 0.6$ & 0.8\\
Southern & CO(1-0) & 0.31 & $7096.9\pm 16.1$ & $178.0\pm 29.2$ & 1.37 & $0.26\pm 0.04$ & $3.1\pm 0.5$ & 1.4\\
2MASX    & CO(1-0) & 1.3  & $7044.1\pm 1.7$  & $150.3\pm 4.0$  & 28.2 & $4.5\pm 0.1$ & 53.7$\pm 1.2$ & 24.2\\
2MASX    & CO(2-1) & 3.6  & $7046.3\pm 2.0$  & $123.8\pm 4.3$  & 61.9 & $8.2\pm 0.3$ & 24.5$\pm 0.9$ & 13.5\\
\hline
\end{tabular}
\tablefoot{
The rms sensitivity values are given for 52~km\,s$^{-1}$ channels in the OC regions and 20~km\,s$^{-1}$ channels in the 2MASX galaxy.}
\end{table*}

\section{Results}
We detected CO(1-0) emission in both the southern and northern regions. Their CO(1-0) spectra, smoothed to 52~km\,s$^{-1}$ resolution, are shown in the right panels of Fig.~\ref{FigOC}. Low rms values of about 0.31~mK and 0.46~mK, respectively, were achieved. In the southern region covering the H$\alpha$ peak, the CO(1-0) line is detected with a ${\rm S/N}= T_{\rm peak}/{\sigma_{\rm rms}}\sim 4.4$, and an integrated ${\rm S/N}= I_{\rm CO}/\sigma_{\rm int}\sim 8.7$, where $\sigma_{\rm int}= {\rm FWHM}\,\sigma_{\rm rms}\sqrt{\Delta \upsilon_{\rm ch}/{\rm FWHM}}$ corresponds to the noise over the spectral channels (of width $\Delta \upsilon_{\rm ch}$) covered by the line. In Fig.~\ref{FigOC} (right panel), the CO(1-0) spectral line is fitted with a Gaussian profile with a central velocity of $\sim 7096$~km\,s$^{-1}$ and a FWHM of 178~km\,s$^{-1}$ (see also Table~1). The integrated intensity is $I_{\rm CO}= 0.26\pm 0.04$~K\,km\,s$^{-1}$, which corresponds to a mean molecular gas column density of $\sim 1.2~M_\odot$\,pc$^{-2}$ and a molecular gas mass of $\sim 1.4\times 10^8~M_\odot$, following the standard relation of \citet{solomon2005}, and assuming a Galactic CO-to-H$_2$ conversion factor of $2\times 10^{20}$~cm$^{-2}$(K\,km\,s$^{-1}$)$^{-1}$. 

Quite unexpectedly, CO(1-0) was also detected in the northern region pointed over the neighboring X-ray peak, which is mostly devoid of H$\alpha$ emission. It is a weaker detection than in the south with a S/N $\sim 3.4$ and an integrated S/N $\sim 4.6$ (see Fig.~\ref{FigOC}, right, top panel). The line is centered at a velocity about 50~km\,s$^{-1}$ larger than the southern line, and its FWHM is smaller by a factor of two. The parameters of a Gaussian fit are given in Table~1. The corresponding mean H$_2$ column density and H$_2$ mass are $\sim 0.6~M_\odot$\,pc$^{-2}$ and $\sim 0.8\times 10^8~M_\odot$, respectively. 

There is no significant line detection in CO(2-1) in either region, although a weak signal at velocities around the CO(1-0) line suggests a marginal detection in the southern region (dashed-line spectra in Fig.~\ref{FigOC}, right panels). The rms sensitivity of the CO(2-1) spectrum per 52~km\,s$^{-1}$ channel is 0.86~mK and 1.1~mK in the southern and northern regions, respectively. Using the widths of the CO(1-0) line, we can estimate the $3\sigma$ upper limit of the CO(2-1) luminosity. The corresponding upper limit on the $L'_{\rm CO(2-1)}/L'_{\rm CO(1-0)}$ line ratio in the southern region is $\sim 0.22$. 
When the size of the CO(1-0) beam is accounted for, which is four-times larger than the CO(2-1) beam, this value may be consistent with typical values of $\sim 0.8$ in normal star-forming galaxies \citep{saintonge2017, leroy2009}. The lower value could also indicate that the molecular gas is diffuse, warmer, and weak in CO emission, due to heating by the neighboring hot gas components \citep{penaloza2017}. In the northern region, most of the CO emission is probably not coming from the center of the beam, but is more widespread within the CO(1-0) beam.

A Gaussian fit to the southern CO(1-0) line leads to a velocity dispersion of $\sim 74$~km\,s$^{-1}$. Such a high value points to multiple molecular clouds with typical velocity dispersion less than a few km\,s$^{-1}$, but it may also be due to a velocity gradient in the beam. The CO velocity dispersion mirrors that of the median value of $\sim 60$~km\,s$^{-1}$ of the H$\alpha$ emission. The velocity dispersion of the northern CO(1-0) emission is $\sim 37$~km\,s$^{-1}$, which may suggest that fewer molecular clouds were covered by the beam. The mean H$\alpha$ velocity dispersion in the northern region is $\sim 52$~km\,s$^{-1}$. Higher resolution CO observations, resolving the emission in these areas, will be required to reveal the detailed morphology and kinematics of CO associated with the OC.

\section{Discussion}
\subsection{Correlation of gas phases}
The detected CO emission has velocities redshifted relative to the H$\alpha$ component. Moreover, in the northern region, the CO emission is shifted to higher velocities relative to the southern region, while the underlying H$\alpha$ emission in the two regions shows the opposite behavior. 

The brightest H$\alpha$ regions from MUSE \citep{ge2021} covered by the southern CO(1-0) beam have a median velocity of $cz\sim 7223$~km\,s$^{-1}$, while the central velocity of the CO(1-0) line is redshifted by $\sim 45$~km\,s$^{-1}$. The relatively large width of the CO line encompasses the H$\alpha$ velocities. In the northern region, the only H$\alpha$ emission is from the filaments at the edges of the CO(1-0) beam, with median velocities $cz\sim 7118- 7230$~km\,s$^{-1}$ (with a representative value of $cz\sim 7182$~km\,s$^{-1}$), thus mostly blueshifted relative to the southern region. However, the northern CO line peaks $\sim 133$~km\,s$^{-1}$ higher and is thus rather strongly redshifted, both relative to the northern H$\alpha$ emission as well as the southern CO line. 

The velocity difference indicates that in the northern region, the cold and warm components are kinematically separated. In the southern region, the CO emission is more kinematically linked to the H$\alpha$ emission, possibly due to a richer distribution of gas that covers a wider range in velocities. 

\begin{figure}
  \centering
  \includegraphics[width=0.4\textwidth]{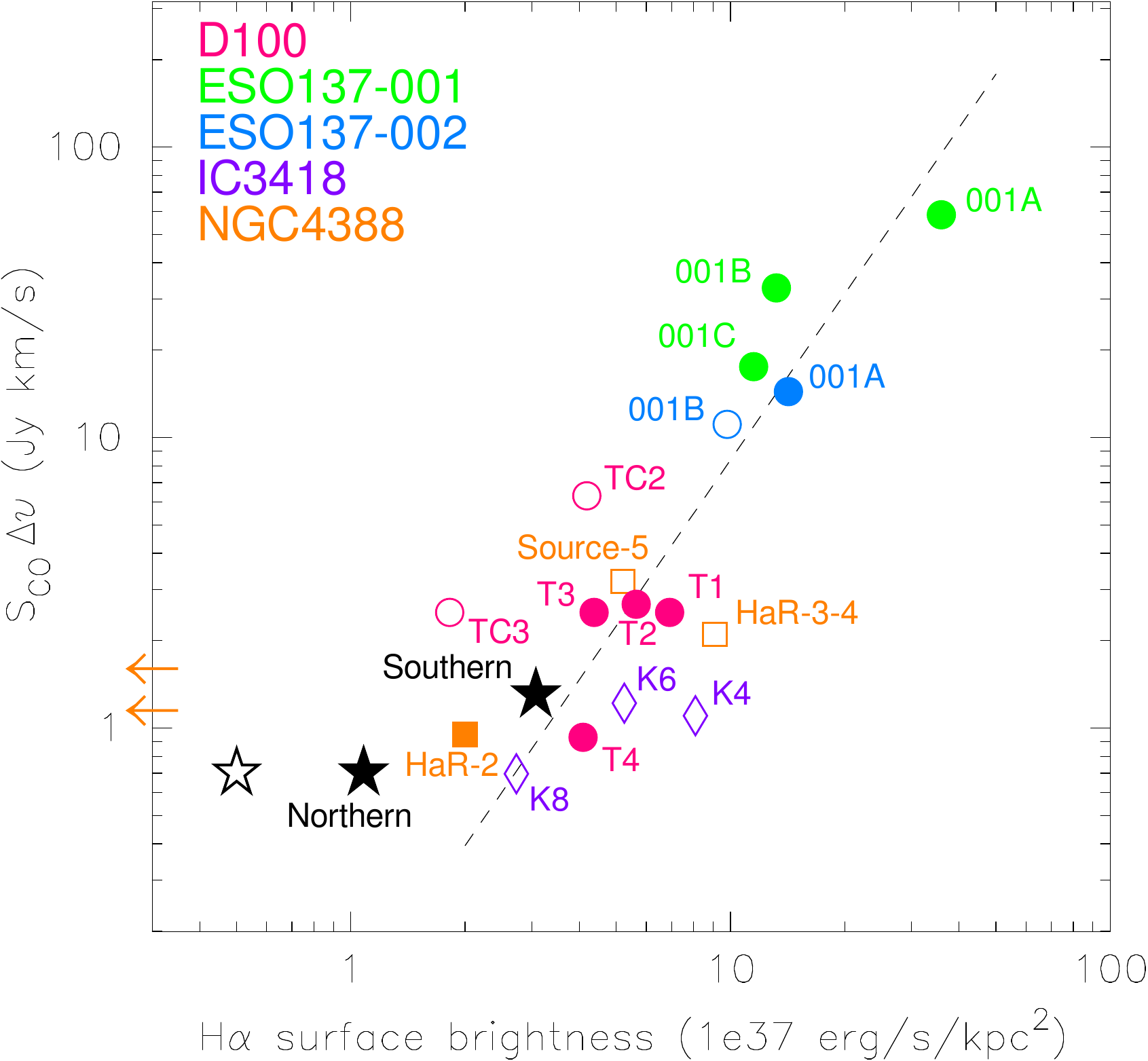}
  \caption{
CO integrated flux vs. H$\alpha$ surface brightness measured in a sample of ram pressure stripped gas tails \citep[][their Fig.~9]{jachym2017}. The location of the two OC regions is shown with filled stars. The open star symbol corresponds to the northern region surface brightness, when the full area of the CO(1-0) beam is used instead of the actual H$\alpha$-emitting area within the beam (filled star).  
}
  \label{FigCO}
\end{figure}

The correlation of cold molecular and warm ionized phases is further explored in Fig.~\ref{FigCO}, where the two regions are plotted on the CO--H$\alpha$ correlation suggested from previous observations of RPS tails \citep{jachym2017}. In the southern OC region, the integrated CO intensity and the H$\alpha$ surface brightness enclosed within the IRAM beam are consistent with the correlation. The northern region is slightly offset -- it has a factor of $\sim 2$ lower H$\alpha$ surface brightness for its CO integrated flux density. The plot in Fig.~\ref{FigCO} comprises tail regions with a predominantly diffuse warm ionized gas component and no or only a few compact HII regions (the highest contribution of HII regions is in the inner part of the tail of the Norma galaxy ESO~137-001 -- point 001A in the plot, where it makes up about 37 \% of the H$\alpha$ flux). The correlation was measured at spatial scales of $\sim 5-10$~kpc. 

In addition, \citet{ge2021} examined the correlation between the warm and hot ionized phases in the OC and compared it with the tight H$\alpha$--X-ray correlation recently found for RPS tails \citep{sun2021}. They found excellent agreement for the southern region, while the X-ray/H$\alpha$ ratio in the region encompassing our northern region is higher by a factor of $\sim 4$ \citep[][see their Fig.~8]{ge2021}.

Thus, the conditions in the southern region covering the main H$\alpha$ cloud seem to be close to the stripped gas in typical RPS tails, where tight H$\alpha$--CO and H$\alpha$--X-ray correlations are observed. In the northern region, on the other hand, the gas may be locally in a different evolutionary phase with the elevated X-ray/H$\alpha$ and CO/H$\alpha$ ratios. In the north, the gas phases are physically separated: the peak of X-ray emission is surrounded by the colder H$\alpha$ emission filaments, from which the CO emission is kinematically offset. It may be expected that the cold gas is also spatially offset from the hot gas, given the surprising fact that there is a cold gas associated with the location of the X-ray peak. 

Spatial and kinematic offsets are characteristic for RPS tails, where different gas components may be both spatially and kinematically offset due to the effects of differential acceleration by ram pressure. For example, offsets of the CO, H$\alpha$, and X-ray emission were observed in the tail of the Norma cluster galaxy ESO~137-001 \citep{jachym2019}. In the Coma cluster galaxy D100, a kinematical offset of $\sim 50$~km\,s$^{-1}$ was measured along the tail between the H$\alpha$- and CO-emitting components \citep{jachym2017}. Future detailed CO mapping will show whether cold and warm gas velocity gradients occur along a consistent direction, and how the X-ray--H$\alpha$ offset relates to it.

\subsection{Mass budget}
Following the estimates of \citet{ge2021}, the mass budget of the part of the OC covered by the MUSE observations is as follows: there is about $8\times 10^7~M_\odot$ of warm H$\alpha$ gas and about $3\times 10^9~M_\odot$ of hot X-ray gas (which is about a third of the total OC X-ray emission). With the $\sim 2.2\times 10^8~M_\odot$ of molecular gas detected in the two regions, our observations indicate that a molecular component may form about 10\% of the total mass of the cloud, and that there is more cold than warm gas. From VLA observations \citep{scott2018}, an upper limit on HI content in the area covering the cloud can be calculated (from their field D observed in the velocity range of $7091- 7463$~km\,s$^{-1}$). For the OC region in the primary beam-corrected cube smoothed spatially to $45''$, which is close to the size of the OC, the measured rms is $\sim 0.67$~mJy in $11$~km\,s$^{-1}$ channels. Assuming the linewidths of the CO lines of 90 and 180~km\,s$^{-1}$, the corresponding HI upper limits (corresponding to a $3\sigma$-significance across three channels) are $\sim 2.2\times 10^8~M_\odot$ and $\sim 3.1\times 10^8~M_\odot$, respectively, thus similar to the lower limit on the molecular gas content.

\subsection{Inefficient star formation}
Analyses by \citet{yagi2017} and \citet{ge2021} indicated that star formation is almost absent in the OC. The upper limit for the star formation rate (SFR) of $6\times 10^{-4}~M_\odot$\,yr$^{-1}$ was estimated from the lack of any GALEX sources in the MUSE field, with the calibration from \citet{kennicutt2012}. The only HII candidate in the MUSE field was identified outside the main H$\alpha$ cloud at RA 11:44:21.4, Dec. +20:10:14.6 \citep[see Fig.~3 in][]{ge2021}. 

For the typical molecular gas depletion time in nearby spiral galaxies $\tau_{\rm dep, H_2}\approx 2$~Gyr \citep[e.g.,][]{bigiel2011, leroy2013}, the molecular gas mass of $2.2\times 10^8~M_\odot$ would correspond to a ${\rm SFR}= M_{\rm H_2}/ \tau_{\rm dep, H_2}\sim 0.1~M_\odot$\,yr$^{-1}$. The observed upper limit on the SFR is thus more than two orders of magnitude lower, indicating that the process of converting molecular gas into stars in the OC is extremely inefficient\footnote{We note that uncertainty in the CO-to-H$_2$ conversion factor cannot account for such a low efficiency and extremely long depletion times.}. The inefficiency further increases when the atomic HI content's upper limit is taken into account. 

The detected molecular gas is in a non-star-forming state. The physical conditions in the OC are likely distinct from galactic disks where star formation typically occurs. Some factors which, under typical conditions, drive and regulate star formation might be absent in the OC (e.g., the gravitational potential of a galaxy disk, stellar feedback, and UV background radiation), while some others are likely present (e.g., heat conduction or shocks from the surrounding ICM). This is similar to the conditions in RPS tails, where star formation can occur when it is still close to the galaxy, or just recently stripped, but it becomes less likely with increasing time and distance.

The CO velocity dispersion in the OC is rather high and similar to the velocity dispersion in the warm gas, which makes the gas less likely to be self-gravitating. While a highly resolved observations of the velocity field would be needed, shocks induced by ram pressure may increase the turbulent kinetic energy which stabilize molecular clouds, increasing their virial parameter and preventing them from collapse \citep[e.g.,][]{sivanandam2010}. 

The density of the molecular gas may play an important role in the low star-formation efficiency. The low mean surface density of $\sim 1~M_\odot$\,pc$^{-2}$ corresponding to the measured integrated CO intensity suggests that the volume density of the molecular gas is very low, or that the fraction of the dense molecular phase that is more tightly correlated with star formation \citep[e.g.,][]{gao2004} in the OC is limited. Recent observations of RPS tails indicated that an important fraction of the molecular component is extended, and its fraction increases with the distance from the parent galaxy \citep{jachym2019, moretti2020}. 

For comparison, the known examples of jellyfish tails with a detected molecular component revealed molecular gas depletion times $> 10^{10}$~yr \citep{jachym2014, jachym2017, moretti2018, moretti2020}. Their star formation rates vary in the range of $\sim 0.01- 1~M_\odot$\,yr$^{-1}$ \citep[e.g.,][]{poggianti2019, cramer2019, kenney2014}.  One of the extreme cases is the tail of the Coma cluster RPS galaxy D100, where for the detected $M_{\rm H_2}\sim 10^9~M_\odot$, the SFR is only $\sim 6.0\times 10^{-3}~M_\odot$\,yr$^{-1}$, and thus $\tau_{\rm dep, H_2}\sim 1.6\times 10^{11}$~yr. 
Long depletion times were previously observed locally in some tidally disturbed galaxies \citep[e.g.,][]{tomicic2018}, gas disks in early-type galaxies that are more stable against fragmentation to dense clumps due to the lower disk self-gravity and increased shear \citep[e.g.,][]{martig2013}, HI-excess galaxies in low-density environments \citep{gereb2018}, or outer parts of spiral disks which, however, are dominated by a (low column density) HI component \citep{bigiel2010, yildiz2017}.

\subsection{Origin of the molecular OC}
Given the large projected distance of $>80$~kpc of the OC from any possible parent galaxy, it is reasonable to suggest that the OC is in an advanced evolutionary state and that the original cold gas was heated due to mixing with the hot ICM. In RPS tails, compact molecular gas detected at large (several tens of kiloparsecs) distances from the main galaxies are expected to have formed in situ from the stripped gas \citep{jachym2019, moretti2020}. 

Dust survival is an essential element of the formation of molecular gas \citep[e.g.,][]{hollenbach1979}. This implies that in the OC, the stripped gas that has mixed with the ICM wind still keeps a fraction of its initial dust content. Recent 3D hydrodynamic simulations of the evolution of cold ($10^3$~K) dusty clouds in a hot wind indicate that if the characteristic cooling timescales of the mixed gas are shorter than the cloud crushing time, the atomic phase can survive, and once the cloud is entrained by the wind, it will cool down and molecular gas may reform \citep{farber2021, gronke2020, kanjilal2021}. Another factor that could help the recent formation of molecular gas is compression due to ram pressure. The H$\alpha$ OC has a velocity gradient nearly aligned in the east-west direction \citep{ge2021}, which, together with the elevated values of the H$\alpha$ velocity dispersion at the west side of the OC, may suggest that the west edge experiences an ICM ram pressure.

\subsection{Molecular content of a nearby RPS galaxy}
About 100~kpc southward of the OC at the same redshift lies the galaxy 2MASX~J11443212+2006238, which has an 85~kpc long H$\alpha$ ram pressure stripped tail \citep[][see Fig.~\ref{FigOC}, left panel]{gavazzi2017}. Both CO(1-0) and CO(2-1) emission was strongly detected in the main body of the galaxy (see the inset spectra in Fig.~\ref{FigOC}, left panel), with a corresponding H$_2$ mass of $\sim 2.4\times 10^9~M_\odot$. Table~\ref{table:1} summarizes the rms sensitivity of the CO data in 20~km\,s$^{-1}$ channels and the parameters of the Gaussian fits. The fits indicate that the line profiles are slightly asymmetric, especially in CO(1-0), likely corresponding to the redistribution of the molecular gas due to ram pressure stripping. The CO emission shows no offset in radial velocity from H$\alpha$ \citep{gavazzi2017}.

2MASX~J11443212+2006238 illustrates an early phase of ram pressure stripping when the tail is connected to a star-forming, H$\alpha$-bright disk \citep{yagi2017}. This is in obvious contrast to the late evolutionary phase of the neighboring OC which has detached from its parent galaxy and has evolved and mixed with the hot ICM. 

Despite the similar velocities and the projected proximity, there is no direct observational evidence that the galaxy is related to the OC; also, 2MASX could not be the RPS parent of the OC either since its radial velocity would likely need to be substantially larger (redshifted) than that of the OC. While the origin of the OC will be studied elsewhere, we note that there are several galaxy pairs with a known HI-deficient late-type galaxy member near the OC \citep{scott2010, scott2018}. A pair tidal interaction loosening the ISM from the pair's gravitational potential with subsequent RPS could also be a viable scenario for the origin of the gas in the OC.

\section{Conclusions}
Using the IRAM 30m telescope, we have detected CO(1-0) emission in two regions of the multiphase OC. This recently discovered object lies at $\sim 800$~kpc north of the center of the Abell 1367 galaxy cluster. About $1.4\times 10^8~M_\odot$ of H$_2$ was detected in the region covering the peak in H$\alpha$ emission (assuming a standard Galactic CO-to-H$_2$ conversion factor) and about $8\times 10^7~M_\odot$ in the region separated by $\sim 12$~kpc, where X-ray emission has a peak, surrounded by H$\alpha$ filaments, as shown by the MUSE observations \citep[see Fig.~1;][]{ge2021}. 

In the region covering the H$\alpha$ peak, the conditions seem to be consistent with that of gas in typical ram pressure stripped tails and follows the established tight H$\alpha$--X-ray and H$\alpha$--CO correlations. In the region covering the X-ray peak, the X-ray/H$\alpha$ and CO/H$\alpha$ ratios are elevated by factors of 4 and 2, respectively, and the mean velocity of the CO emission is offset from the warm ionized component by more than 100~km\,s$^{-1}$ . This may support the scenario of the RPS origin of the OC since spatial and kinematics offsets of gas phases (as well as the presence of a velocity gradient) are characteristic of RPS tails. 

The CO velocity dispersion in the two regions is rather high ($\sim 70$ \& 40~km\,s$^{-1}$). Together with the presumably low molecular gas volume density, this may be the principal factor in the extremely low star formation efficiency of the detected molecular gas ($\tau_{\rm dep, H_2}\sim 3.7\times 10^{11}$~yr). 

The OC likely represents a late evolutionary product of a gas stripping event of an unidentified cluster galaxy. Its total mass is dominated by the hot ionized component, while the molecular gas forms about 10\% of the total mass budget, similar to the upper limit contribution by HI emission calculated from existing VLA observations.

In contrast to the aged OC, we also observed the nearby galaxy 2MASX~J11443212+2006238, which is in an early stage of ram pressure stripping, and detected $\sim 2.4\times 10^9~M_\odot$ of molecular gas in its main body.

\begin{acknowledgements}
We warmly thank the IRAM 30-m observatory staff for their help during the observations and with the preparation of observing scripts. 
PJ acknowledges support from the project LM2018106 of the Ministry of Education, Youth and Sports of the Czech Republic and from the project RVO:67985815.
MS acknowledges support provided by the NASA grant 80NSSC19K0953 and the NSF grant 1714764.
TS acknowledges support by Funda\c{c}\~{a}o para a Ci\^{e}ncia e a Tecnologia (FCT) through national funds (UID/FIS/04434/2013), FCT/MCTES through national funds (PIDDAC) by this grant UID/FIS/04434/2019 and by FEDER through COMPETE2020 (POCI--01--0145--FEDER--007672). TS also acknowledges support from DL 57/2016/CP1364/CT0009.
\end{acknowledgements}

\bibliographystyle{aa}
\bibliography{OCacc}

\end{document}